\documentclass[%
aip,
 sd,
 amsmath,amssymb,
 reprint,%
]{revtex4-2} % Template provided by structural dynamics on Overleaf

\usepackage[english]{babel}% 
\usepackage{graphicx}% Include figure files
\usepackage{dcolumn}% Align table columns on decimal point
\usepackage{bm}% bold math
\usepackage{textcomp}
\usepackage[utf8]{inputenc}
\usepackage[T1]{fontenc}
\usepackage{mathptmx}

\usepackage{hyperref}
\usepackage{xcolor}
\definecolor{linkblue}{RGB}{43,168,224}
\hypersetup{linkcolor  = linkblue,citecolor  = linkblue,urlcolor   = linkblue,colorlinks = true,}

\begin{document}

\preprint{AIP/123-QED}

\title{Phonon-dominated energy transport in purely metallic heterostructures}
%\title{Phonon-dominated energy transport in purely metallic heterostructures at the nanoscale}

\author{M.~Herzog}
%\email{marc.herzog@uni-potsdam.de}
\affiliation{Institut f\"ur Physik \& Astronomie,  Universit\"at Potsdam, Karl-Liebknecht-Str. 24-25, 14476 Potsdam,  Germany}

\author{A.~von~Reppert}
\affiliation{Institut f\"ur Physik \& Astronomie,  Universit\"at Potsdam, Karl-Liebknecht-Str. 24-25, 14476 Potsdam,  Germany}

\author{J.-E.~Pudell}
\affiliation{Institut f\"ur Physik \& Astronomie,  Universit\"at Potsdam, Karl-Liebknecht-Str. 24-25, 14476 Potsdam,  Germany}
\affiliation{Helmholtz Zentrum Berlin, Albert-Einstein-Str.~15, 12489 Berlin, Germany}

\author{C.~Henkel}
%\email{marc.herzog@uni-potsdam.de}
\affiliation{Institut f\"ur Physik \& Astronomie,  Universit\"at Potsdam, Karl-Liebknecht-Str. 24-25, 14476 Potsdam,  Germany}

\author{M.~Kronseder}
\affiliation{Institut f\"ur Experimentelle und Angewandte Physik, Universit\"at Regensburg, Germany}

\author{C.~H.~Back}
\affiliation{Institut f\"ur Experimentelle und Angewandte Physik, Universit\"at Regensburg, Germany}
\affiliation{Physics Department, Technical University Munich, 85748 Garching, Germany}

\author{A.~Maznev}
\affiliation{Department of Chemistry, Massachusetts Institute of Technology, Cambridge, MA 02139, USA}

\author{M.~Bargheer}
\email{bargheer@uni-potsdam.de}
\affiliation{Institut  f\"ur Physik \& Astronomie, Universit\"at Potsdam,  Karl-Liebknecht-Str. 24-25, 14476 Potsdam, Germany}
\affiliation{Helmholtz Zentrum Berlin, Albert-Einstein-Str.~15, 12489 Berlin, Germany}

%\homepage{http://www.udkm.physik.uni-potsdam.de} 
\date{\today}

\begin{abstract}
We use ultrafast x-ray diffraction  to quantify the transport of energy in laser-excited nanoscale Au/Ni bilayers. Electron transport and efficient electron-phonon coupling in Ni convert the laser-deposited energy in the conduction electrons within a few picoseconds into a strong non-equilibrium between hot Ni and cold Au phonons at the bilayer interface. Modeling of the subsequent equilibration dynamics within various two-temperature models confirms that for ultrathin Au films the thermal transport is dominated by phonons instead of conduction electrons because of the weak electron-phonon coupling in Au.% of nanoscopic and the concomitant gradient in lattice temperatures.
%We present ultrafast x-ray diffraction (UXRD) experiments on laser-excited thin Au-Ni bilayers on MgO substrates in order to investigate the role of electrons and phonons for thermal transport and equilibration between the two metals. The excitation of the metal bilayer generates hot electrons and the excess energy is quickly dumped into the Ni lattice while the Au lattice is only very slowly heated on a time scale of tens of picoseconds owing to the large discrepancy in electron-phonon (e-ph) coupling strength of Ni and Au. This long-lasting large temperature gradient between the Au and Ni lattice causes a significant thermal transport by phonons between the two metals which is typically small or simply disregarded. We use a simplified two-temperature model (2TM) to qualitatively explain our findings in a comprehensive way and we present numerical calculations from a diffusive 2TM that reproduce the UXRD data very well and which evidence that phonons even dominate the thermal transport across the metal-metal interface as the Au layer thickness approaches the single-digit nanometer scale.
\end{abstract}

\maketitle

\section{\label{sec:intro}Introduction} 
Thermal transport in metallic heterostructures at the nanoscale is highly relevant for diverse fields ranging from thermal management in nanoelectronics\cite{pop2010,oomm2021}, spin-caloritronics\cite{baue2012} and ultrafast spin dynamics\cite{agar2021} to photothermal processes \cite{baff2013} and plasmonic chemistry\cite{zhan2018}. The thermal transport in elemental metals is usually dominated by conduction electrons as the thermal conductivity by phonons (lattice vibrations) is typically one or two orders of magnitude smaller than its electronic counterpart\cite{cahi2014,wang2016}. In thermal equilibrium the mean-free path and thus the thermal conductivity of electrons is typically limited by the electron-phonon interaction which simultaneously excites phonons. Under non-equilibrium conditions, for example, in metals right after absorption of an ultrashort laser pulse, the enhanced specific heat and thermal conductivity of the hot electron gas in conjunction with ballistic or superdiffusive electron transport tend to even enhance the role of electron thermal transport\cite{cork1988,hohl2000,batt2012,pude2020b}.

Modern device architectures as well as samples used in material science are typically nano-scale heterostructures containing materials with different properties to enable specific functionalities. The importance of phonon thermal transport is obvious in metal-insulator heterostructures. Thermal resistance network models assume a dominant role of the phonons for heat transport through sub 10 nm metal layers of Mo/Si superlattices
\cite{li2012}. Phonon thermal transport in metal-metal heterostructures has been rarely studied as it is often disregarded for being a marginal effect\cite{gund2005,wang2012,oomm2021}. 

A standard experimental method in order to address dynamical processes on picosecond time scales is the pump-probe technique. Here, the sample of interest is driven out of equilibrium by an intense laser pump pulse and the resulting changes in the multipartite and mutually coupled degrees of freedom are detected with a delayed probe pulse. The characteristics of the probe pulse depend on the desired information. Using a visible or near-infrared probe pulse provides access to the coupled dynamics of electrons, phonons and spins in condensed matter, e.g., by means of thermoreflectance, transient absorption and the magneto-optical Kerr effect (MOKE) \cite{wang2012,choi2014b,delF2000,jang2020}. Although tremendous insight in the involved physical processes has been gained in the last decades exploiting these techniques, the experimental results are often difficult to interpret, in particular, far from non-equilibrium conditions or in heterostructures with dimensions approaching the skin depth of the probe light\cite{jang2020}. Experimental techniques that are able to precisely determine the layer- and subsystem-specific energy content during the non-equilibrium dynamics in nanoscopic metal heterostructures are sparse. This has limited the dimensions of the investigated metal films in previous experiments to several tens of nanometers. Under such circumstances the phonon contribution to the energy exchange among the constituents was indeed negligible or very small\cite{gund2005,wang2012}.

Here we showcase that the exchange of energy between nanoscale metal films can be dominated by the phonons instead of the conduction electrons. We exemplarily consider Au-Ni bilayers supported by an insulating MgO substrate. After excitation with an ultrashort laser pulse, we exploit the material-specific sensitivity of ultrashort x-ray diffraction (UXRD) to infer the evolution of the energy content from the strain response in each of the metal layers. The time-dependent shifts of the layer-specific Bragg peaks precisely quantify the energy exchange between the ultrathin films. We model the underlying dynamics using two-temperature models with various levels of complexity for the coupled electron-phonon systems in the heterostructure. As pictorially summarized in Fig.~\ref{fig:UXRDdata}(a), the analysis of our experimental data reveals the intriguing scenario that in few-nm thin films
energy transport across the metal-metal interface by phonons (red arrows) dominates over the transport via electrons and electron-phonon coupling (blue arrows). The crucial ingredient for this unconventional thermal transport between the two metals is their strongly differing electron-phonon coupling (EPC) strength. The very efficient EPC in Ni rapidly concentrates almost the entire laser-deposited energy in the Ni phonons thus generating a large gradient in the phonon temperature. Although the conduction electrons have a superior thermal conductivity, the weak EPC represents a particularly tight bottleneck in very thin Au films for transferring energy into the large heat reservoir of the Au phonons. This leaves the major contribution to the thermal interlayer equilibration to phonon transport.
%UXRD is thus a perfectly suited tool to explore nanoscale thermal transport down to structure sizes of few unit cells where quantum effects are expected to become increasingly important.

\section{Experiment} \label{sec:introCons}
%\begin{figure}[tb]
%  \centering
  %\includegraphics[width = \columnwidth]{figures/schematic_system.png}
%  \includegraphics[width = %\columnwidth]{figures/fig1new.eps}
%  \columnwidth]{figures/Fig1MB.jpg}
%  \caption{Schematic of the investigated Au-Ni heterostructure with variable Au thickness on a MgO substrate. The figure visualizes the energy flow inside the heterostructure after the Au and Ni electrons have equilibrated with the Ni lattice. The arrows depict the energy flow via phonons (red)  and electron-phonon coupling (EPC) in combination with electron transport (blue), respectively. The transport via electrons is very efficient and long-ranging, however, the weak EPC in Au forms a bottleneck for this channel. Therefore, on short length scales, the efficiency of phonon transport can be superior.}
%  \label{fig:sketch}
%\end{figure}
%Fig.~\ref{fig:sketch} sketches the optical excitation of the electronic system of Au/Ni heterostructures which is coupled to the phonon system of the heterostructure and the substrate.
%In order to investigate the role of thermal transport by phonons in thin metal heterostructures, 
A series of Au-Ni bilayers of various thickness combinations was grown on commercial MgO substrates by molecular beam epitaxy (MBE). The thickness of the Au and Ni layers (see Table~\ref{tab:samples}) were precisely characterized by x-ray reflectivity (XRR) measurements at the KMC3-XPP beamline at BESSY II\cite{roes2021}. Figure~\ref{fig:results}(a) schematically shows the layered structure of the samples S1-S4.

\begin{figure}[t]
  \centering
  \includegraphics[width = %\columnwidth]{figures/fig2_sample12.eps
  \columnwidth]{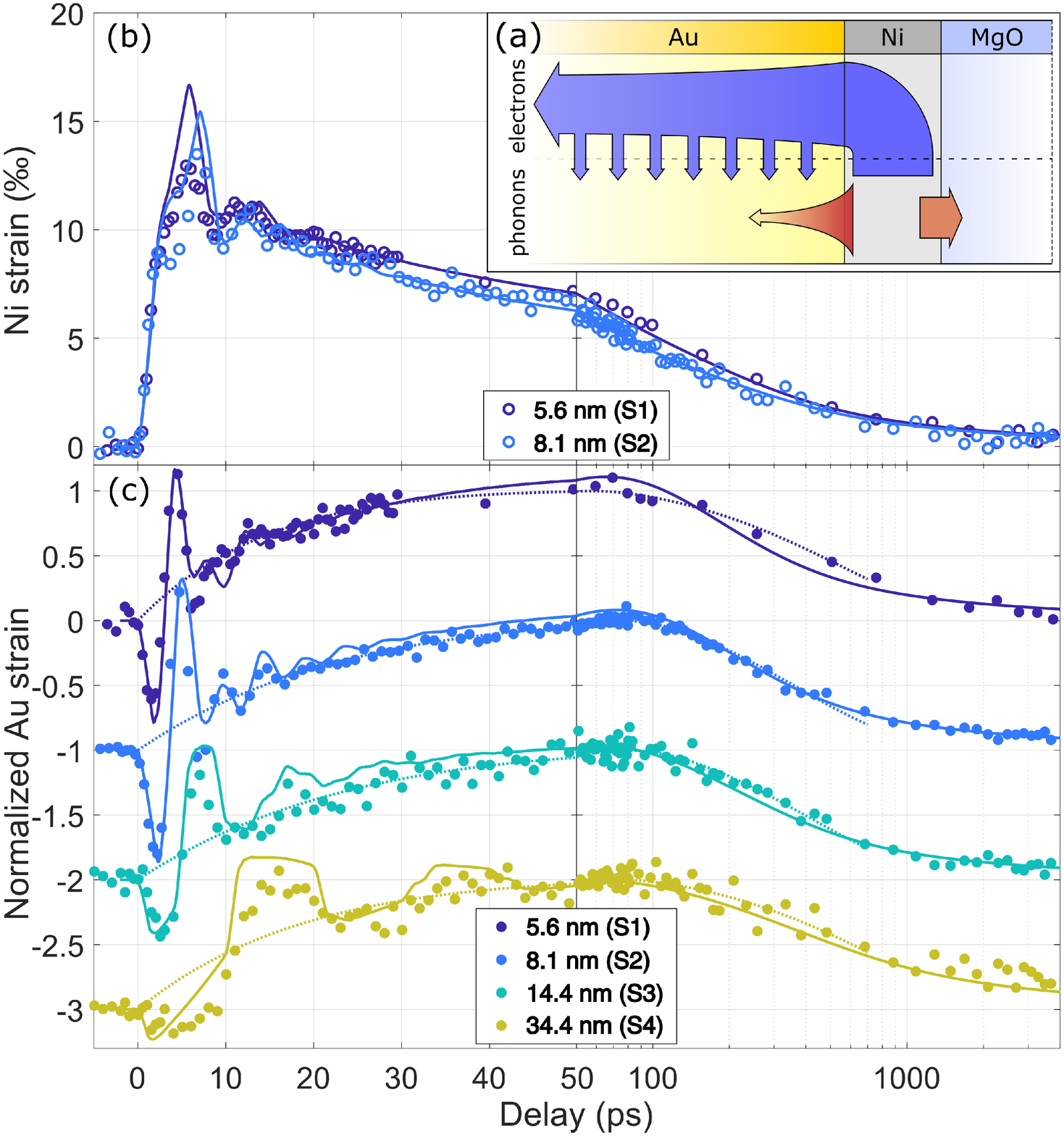}
  \caption{(a) Schematic of the Au-Ni heterostructure with variable Au thickness on a MgO substrate. The figure visualizes the energy flow inside the heterostructure after the Au and Ni electrons have equilibrated with the Ni phonon system. The arrows depict the energy flow via electron diffusion and EPC (blue) and via direct phonon diffusion (red). The transport via electrons is very efficient and long-ranging, however, the weak EPC in Au forms a bottleneck for this channel. (b)-(c) Results of UXRD experiments on samples with systematic thickness variation for a common incident pump fluence of approx.~10\,mJ/cm$^2$. Panel (b) and (c) depict the transient Ni and Au strain, respectively, after excitation with an ultrashort laser pulse. The symbols represent the experimentally measured transient layer strains which are derived from the shifts of the respective Bragg reflections in reciprocal space. Dashed lines are empirical fits according to eq.~(\ref{eq:fitFunc}) representing the rise and decay of the incoherent Au stress/strain. The solid lines are results of the numerical modelling of the coherent and incoherent layer strains using the d2TM. Note that the Au strain transients in panel (c) are offset and normalized between 50 and 100\,ps for a better comparability. The legends provide the sample number and the Au thickness.
  }
  \label{fig:UXRDdata}
\end{figure}

We use ultrashort x-ray pulses generated by a table-top laser-driven plasma x-ray source (PXS) to record transient angular shifts of Au (111) and Ni (002) Bragg reflections after excitation of the sample with an ultrashort 800\,nm laser pulse\cite{schi2012,schi2013a}. The shifts of the material-specific Bragg reflections in reciprocal space are converted to a time-dependent average lattice strain for each layer\cite{zeus2021}.
The experimental results for the samples S1-S4 at a common incident pump fluence of approx.~10\,mJ/cm$^2$ are gathered in Fig.~\ref{fig:UXRDdata}.
%Since this stress is linearly related to the energy density inside a material, UXRD serves as a perfect tool to investigate the energy transport in ultrathin heterostructures. \textcolor{gray}{(If this "assumption" is attacked by the referees, we also have very high fluence data from sample 2.5 where the Debye-Waller-like intensity [being a measure of temperature instead of energy] is essentially consistent wth the strain transients!)}
%The results of the UXRD experiments on the Au-Ni bilayer samples series are summarized in .
The transient strain evolution in Ni and Au (Fig.~\ref{fig:UXRDdata}(b) and (c)) consists of a superposition of coherent longitudinal acoustic phonon wavepackets and incoherent contributions (thermal expansion)\cite{pude2018,pude2020b}. Right after excitation at zero time delay the Ni layer rapidly expands\footnote{Note that not all samples exhibited a sufficiently intense Ni Bragg peak for measuring the transient Ni strain with satisfactory statistics.}, evidencing the expected sub-picosecond energy transfer from the laser-heated conduction electrons into Ni \cite{pude2018,pude2020b}. This quick stress rise in the Ni lattice launches a coherent strain wavepacket which causes the oscillatory signatures in the first tens of picoseconds in Fig.~\ref{fig:UXRDdata}(b) and (c) as it is reflected and transmitted through the layer interfaces. All transients reveal that the Au layer is compressed immediately after excitation, which shows that the Au contains substantially less energy than the Ni layer on that time scale. Irrespective of the layer thickness, the incoherent strain within Au grows extraordinarily slowly compared to the intrinsic few-picosecond EPC time scale\cite{well1999,choi2014b} and reaches its maximum around 80\,ps after excitation\cite{pude2018} which is much longer than previously expected\cite{oomm2021}. At later time scales both layers obviously relax by heat diffusion into the MgO substrate.

\section{Modified Two-Temperature Model (m2TM)} \label{sec:Modelling}
The pump pulse energy is absorbed exclusively by the conduction electrons inside the Au-Ni bilayer whereby the electron-phonon (e-ph) system of the metals is strongly driven out of equilibrium. % which requires the consideration of separate thermal reservoirs representing conduction electrons and the lattice as sketched in Fig.~\ref{fig:sketch}.
Immediately after laser excitation neither the electron nor the phonon system are perfectly described by Fermi and Bose distribution functions, respectively\cite{mald2020}. 
However, our experimental observable, the strain, is sensitive to the local energy densities $\Delta \rho^Q(z,t)$ rather than to details of the distribution functions\cite{barr1957,barr1980,matt2022a}. Therefore, we define local (quasi-equilibrium) temperatures $T_\text{e/ph}(z,t)$ for each subsystem via $\Delta\rho^Q_\text{e/ph}=\int_{\Delta T_\text{e/ph}} C_\text{e/ph}(T_\text{e/ph})\,\text{d}T_\text{e/ph}$. 

The main idea of this paper is that the non-equilibrium between electrons and phonons in the metal heterostructure is converted into a spatial temperature step, where the Ni phonons are hot and the Au phonons are cold. This temperature gradient in the phonon system efficiently drives thermal transport by phonons across the interface. The alternative channel for heat transport via the conduction electrons and EPC is proportional to the Au layer thickness. That is, for very thin Au films the EPC is inefficient and the equilibration of the phonon systems across the interface is thus predominantly mediated by phonons.

%\squeezetable
\begin{table}[t]
\centering
\begin{tabular}{|l| c c c c c |}
\hline
 & S1 & S1.1 & S2 & S3 & S4\\
\hline
 Au thickness (nm) & 5.6  & 5.7  & 8.1  & 14.4 & 34.4 \\
 Ni thickness (nm) & 12.4 & 21.2 & 11.4 & 7.0  & 6.0  \\
\hline
\end{tabular}
\caption{Au and Ni layer thickness in nm in the investigated samples as determined by x-ray reflectivity.}
\label{tab:samples}
\end{table}

Before we discuss a diffusive two-temperature model (d2TM) with proper spatial and temporal resolution\cite{pude2020b}, we emphasize this main outcome by approximating the situation in a modified two-temperature model (m2TM), which describes the stress that drives the lattice strain in ultrathin films on the ps-time scale \cite{pude2018}. It disregards spatial gradients, thus being suited for rather thin films, and separates the phonon temperature in the material with weak EPC (Au) from the common temperature of conduction electrons (Au+Ni) and strongly coupled phonons (Ni).
This is equivalent to assuming the entire electron system of the heterostructure to be in thermal equilibrium with the Ni phonons.
Hence, in the m2TM the transient temperature of the Au phonons $T_\text{Au}^\text{ph}$ and the combined system of Au and Ni electrons and Ni phonons $T_\text{Ni}$ evolve according to:%. The differential equations (DE) of this condensed 2TM then read\cite{pude2018}
\begin{align}
    d_\text{Au} C_\text{Au}^\text{ph} \frac{\partial T_\text{Au}^\text{ph}}{\partial t} &= G_\text{Au-Ni}\left( T_\text{Ni} - T_\text{Au}^\text{ph} \right) \label{eq1:2TM} \\
    d_\text{Ni} C_\text{Ni} \frac{\partial T_\text{Ni}}{\partial t} &= G_\text{Au-Ni}\left( T_\text{Au}^\text{ph} - T_\text{Ni} \right) + G_\text{sub}\left( T_\text{sub} - T_\text{Ni} \right) \label{eq2:2TM}
%    d_\text{Au} C_\text{Au}^\text{l} \frac{\partial T_\text{Au}^\text{l}}{\partial t} &= d_\text{Au} g_\text{Au}\left( T_\text{Ni} - T_\text{Au}^\text{l} \right) \label{eq1:2TM} \\
%    d_\text{Ni} C_\text{Ni} \frac{\partial T_\text{Ni}}{\partial t} &= d_\text{Au} g_\text{Au}\left( T_\text{Au}^\text{l} - T_\text{Ni} \right) \textcolor{red}{+ G_\text{sub}\left( T_\text{sub} - T_\text{Ni} \right)} \label{eq2:2TM}
\end{align}
where $T_\text{sub}$ is the substrate temperature. Due to the one-dimensional nature of the dynamics in homogeneously excited multilayers we formulate the above equations using conductances $G_i$ representing an energy flow per time and cross-sectional area. Moreover, $d_\text{Au/Ni}$ and $C_\text{Au/Ni}$ are the layer thickness and volumetric heat capacity of the Au and Ni layer, respectively. %We neglect the small electron contributions to the specific heat. 

%In the literature, it is usually argued that direct phonon thermal transport in metal heterostructures is negligibly small\cite{gund2005,wang2012}.
\section{Electronic transport alone does not explain the data} 
In a first approach, we disregard direct phonon thermal transport across the Au-Ni interface as is usually done in literature\cite{gund2005,wang2012}. We only account for electron transport and subsequent EPC in Au. In this simplified case the corresponding conductance is given by $G_\text{Au-Ni}=d_\text{Au} g_\text{Au}$\cite{choi2014b}, as long as $d_\text{Au}$ is well below the mean electron diffusion length given by $L_\text{diff}^\text{e}=\sqrt{\kappa_\text{Au}^\text{e}/g_\text{Au}}\approx 125$\,nm\cite{well1999,wang2012}. Here, $g_\text{Au}$ is the EPC parameter in Au. The coupling $G_\text{sub}$ across the Ni-MgO interface is about an order of magnitude weaker, since e-ph equilibration and heat transfer into the MgO substrate appear on well-separated time scales (Fig.~\ref{fig:UXRDdata}(c)). Disregarding the last term in eq.~(\ref{eq2:2TM}) and assuming temperature-independent material properties, we may first solve the m2TM by an exponential rise and fall of the Au and Ni temperatures, respectively,
\begin{equation}
    \Delta T_\text{Ni} \propto e^{-\frac{t}{\tau_\text{rise}}}, \qquad \Delta T_\text{Au}^\text{ph} \propto \left( 1-e^{-\frac{t}{\tau_\text{rise}}} \right) \label{eq:2TMsolution}
\end{equation}
where the rise time of the Au phonon temperature is given by
\begin{align}
    \tau_\text{rise}^{-1} &= G_\text{Au-Ni} \left( \frac{1}{d_\text{Au}C_\text{Au}^\text{ph}}+\frac{1}{d_\text{Ni}C_\text{Ni}} \right)  \label{eq1:tau} \\
    &= g_\text{Au} \left( \frac{1}{C_\text{Au}^\text{ph}}+\frac{d_\text{Au}}{d_\text{Ni}}\frac{1}{C_\text{Ni}} \right). \quad \text{(no phonons)} \label{eq2:tau}
\end{align}
Eq.~(\ref{eq2:tau}) is an approximation for the simplified case of exclusive energy exchange via conduction electrons.
%In the context of the m2TM introduced above [eqs.~(\ref{eq1:2TM}) and (\ref{eq2:2TM})], the rise time of the Au lattice temperature (and thus to a first approximation the incoherent lattice stress and strain) is given by eq.~(\ref{eq1:tau}) while eq.~(\ref{eq2:tau}) can be applied if direct energy exchange between the phonon systems of Au and Ni can be neglected. 
\begin{figure}[t]
  \centering
    \includegraphics[width = 0.9\columnwidth]{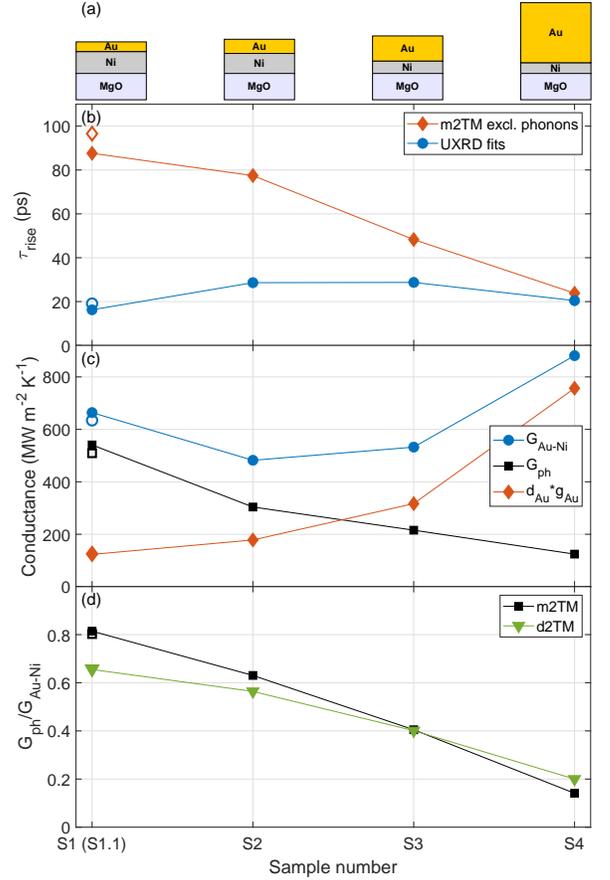}
  \caption{(a) Schematics of the investigated metal heterostructures S1-S4. The empty symbols in the following panels represent a sample S1.1 similar to S1 with identical Au thickness but half the Ni thickness.
  (b) Au strain/stress rise times $\tau_\text{rise}$ derived from the experimental data (blue bullets) and the corresponding prediction from the m2TM in absence of phonon transport (red diamonds) using an EPC constant $g_\text{Au}=2\cdot10^{16}\,$W/(m$^3$\,K).
  (c) Total conductance $G_\text{Au-Ni}$ (blue bullets) determined from the measured $\tau_\text{rise}$ in (b) via eq.~\eqref{eq1:tau}. Red diamonds show the conductance via electrons ($d_\text{Au} g_\text{Au}$) assuming $g_\text{Au}=2\cdot10^{16}\,$W/(m$^3$\,K). Black squares present the difference of the former two which is attributed to the phonon conductance $G_\text{ph}$.
  (d) Relative contribution of phonon transport to the total energy arriving in the Au phonon system as derived from the m2TM (black squares) and the d2TM (green triangles). Both models predict that for Au films below approx. 10\,nm thermal transport via phonons dominates.}
  \label{fig:results}
\end{figure}
On time scales beyond 100\,ps the layers and subsystems have equilibrated and the slowly relaxing strain of both layers indicate that heat diffuses from the bilayer into the substrate via phononic thermal transport. Assuming a constant temperature of the substrate\footnote{This is a reasonable approximation on the sub-nanosecond time scale given the large thermal conductivity of MgO of approx. 50\,W/(m\,K)\cite{hofm2014}.}, eqs.~(\ref{eq1:2TM}) and (\ref{eq2:2TM}) imply an exponential decay of the bilayer temperature. In total, our m2TM suggests a time-dependent temperature of the Au phonons of the form
\begin{equation}
    \Delta T_\text{Au}^\text{ph} \propto \left( 1-e^{-\frac{t}{\tau_\text{rise}}} \right) e^{-\frac{t}{\tau_\text{fall}}}. \label{eq:fitFunc}
\end{equation}
In order to check the validity of the assumption that vibrational thermal transport between the metal layers is negligible in our m2TM, we fit the experimental UXRD transients for the Au layer with eq.~(\ref{eq:fitFunc}) ignoring the oscillatory contributions to the signal. Here we assume that the thermal expansion coefficient of Au is approximately temperature-independent, which is reasonable above its Debye temperature $\Theta_\text{D} \approx 185\;$K\cite{lynn1973}. The best fits to the normalized UXRD data are shown in Fig.~\ref{fig:UXRDdata}(c) as dashed lines. The incoherent rise and decay dynamics of the Au layer strain are well reproduced. The obtained rise times $\tau_\text{rise}$, are shown in Fig.~\ref{fig:results}(b) as blue bullets with only weak variations from sample to sample. In contrast, the rise times predicted by eq.~(\ref{eq2:tau}) of our m2TM shown as red diamonds vary by a factor of five and only agree for the sample with the thickest Au layer.

This observation calls for adding a mechanism to the model which accelerates energy transfer into Au phonons especially in ultrathin Au layers. One option would be a strongly enhanced EPC constant $g_\text{Au}$. Indeed, scattering of the hot conduction electrons at the interface or grain boundaries may couple to surface and interface vibrational modes of the lattice. Also non-thermal electron distributions may alter $g_\text{Au}$. However, for the moderate Au electron temperature increase after equilibration with Ni of only a few hundred Kelvin in the present cases, previous experiments and theory suggest only tiny variations of $g_\text{Au}$\cite{hopk2009,soko2015,soko2017,webe2019} and a required increase by almost one order of magnitude is thus unreasonable. Alternatively, the direct thermal transport between the hot Ni and cold Au via phonons is an inevitable process that must contribute to an accelerated energy transfer from Ni to Au, especially on the sub-10 nm length scale of the phonon mean-free path.\cite{li2012,choi2014b}

It is straightforward to introduce an interface conductance $G_\text{ph}$ into our m2TM representing direct phonon thermal transport across the Au-Ni interface. According to Fig.~\ref{fig:UXRDdata}(a) it acts in parallel to the conductance via %cross-plane electron transport and 
EPC in Au. The coupling constant in eqs.~\eqref{eq1:2TM} and \eqref{eq2:2TM} then reads
\begin{equation}
    G_\text{Au-Ni} = d_\text{Au} g_\text{Au} + G_\text{ph}. \label{eq:cond}
\end{equation}
From the experimental rise times $\tau_\text{rise}$ (Fig.~\ref{fig:results}(b)) we obtain the total conductance $G_\text{Au-Ni}$  via eq.~\eqref{eq1:tau} and then solve  eq.~\eqref{eq:cond} for the phonon interface conductance $G_\text{ph}$, assuming a constant EPC constant $g_\text{Au}=2\cdot10^{16}\,$W/(m$^3$\,K)\cite{hohl2000}. Figure~\ref{fig:results}(c) compares the individual conductances obtained by this procedure. We find an average phonon interface conductance $G_\text{ph}$ on the order of a few 100\,MW/(m$^2$\,K). As a first consistency check, the quantity $G_\text{ph}$ can be estimated for a Au-Ni interface using the respective acoustic phonon dispersions within the radiation limit\cite{swar1989,ston1993}. The calculations yield consistent values on the order of 50-300\,MW/(m$^2$\,K) depending on the assumptions being made with respect to mode conversion at the interface.
%The Au-Ni interface conductance obtained by fitting our UXRD data using the m2TM lie perfectly within the expected range for a Au-Ni interface.
Fig.~\ref{fig:results}(c) directly compares the EPC channel (red diamonds) and the phonon transport across the interface (black squares). For very thin Au films below approx.~10\,nm (S1 \& S2) the phonon interface conductance is larger than the EPC conductance. The main reason for this is that the energy transport into the Au phonons via EPC scales the Au film thickness $d_\text{Au}$ 
and thus becomes very inefficient in ultrathin films. The analysis of our UXRD data presented so far strongly suggests that the direct phonon-mediated energy transport in nanoscopic metal heterostructures cannot generally be ignored and at structure dimensions of a few nanometers the energy transport can even be dominated by phonon transport.

\section{Diffusive Two-Temperature Model (d2TM)} \label{sec:modelling}
The m2TM relies on the very different EPC in Au and Ni, which is the main ingredient for different phonon temperatures in Au and Ni.
%dominant phonon-based energy transport in sufficiently thin metallic bilayers.
In order to cross-check the validity of the m2TM and to resolve the energy flow within Au and Ni layers, we numerically solve a spatially resolved diffusive 2TM (d2TM) including electron and phonon heat diffusion. We utilize our simulation toolbox udkm1Dsim\cite{schi2014,schi2021} which proved to yield precise modelling of experimental UXRD data in the context of non-equilibrium e-ph dynamics and transport in one-dimensional systems\cite{pude2020b,matt2022a}. Although ballistic electron transport is relevant on a sub 100 fs timescale, the diffusive treatment yields useful results in the discussed context, because the equilibration within the electron system is much faster than both the time-resolution of the experiment and any other dynamics in the samples. The coupled differential equations of the d2TM read
\begin{align}
    C_\text{e}(T_\text{e}) \frac{\partial T_\text{e}}{\partial t} &= \frac{\partial}{\partial z} \left( \kappa_\text{e}(T_\text{e},T_\text{ph}) \frac{\partial T_\text{e}}{\partial z} \right) - g(T_\text{e}-T_\text{ph}) + S(z,t) \label{eq1:diff2TM} \\
    \underbrace{C_\text{ph} \frac{\partial T_\text{ph}}{\partial t}}_{\dot\rho^Q_\text{tot,ph}} &= \underbrace{\frac{\partial}{\partial z} \left( \kappa_\text{ph} \frac{\partial T_\text{ph}}{\partial z} \right)}_{\dot\rho^Q_\text{ph}} + \underbrace{g(T_\text{e}-T_\text{ph})}_{\dot\rho^Q_\text{ep}} \label{eq2:diff2TM} %\\
    %&= j_\text{ph}(z,t)+j_\text{e-ph}(z,t) \label{eq:j}
\end{align}
where the upper/lower equation represents the evolution of the electron/phonon temperature after excitation of the electrons via absorption of the laser pulse energy represented by the source term $S(z,t)$. Here, $z$ is the spatial coordinate perpendicular to the sample surface. The quantities $C_\text{e/ph}$, $\kappa_\text{e/ph}$ and $g$ are the specific heat and thermal conductivity of the electrons and the phonons as well as the EPC constant, respectively. As we investigate heterostructures consisting of different materials including their interfaces these quantities have an explicit spatial dependence. 
We also take into account the explicit temperature dependencies $C_\text{e}=\gamma T_\text{e}$ and $\kappa_\text{e}=\kappa_\text{e}^0 T_\text{e}/T_\text{ph}$ implied by the Sommerfeld model of free electrons\cite{pude2020b}, whereas we omit the weak temperature dependence of the phonon counterparts.
We assume negligible interface resistance for the electrons \cite{wang2012}, but in order to parametrize the energy exchange across the metal-metal interface by phonons we include a finite interface conductance $G_\text{ph}$ in terms of a tiny interface layer with suppressed phonon conductivity.
Note that the m2TM fitting analysis presented above correctly recovers the $G_\text{ph}$ applied in a corresponding d2TM calculation given the latter uses the same simplifying assumptions such as infinite conductivities in Au and Ni and vanishing heat conduction into the substrate.
%Without this TBC the calculated phonon transport would be even more pronounced. %by means of a strongly reduced thermal conductivity at the Au-Ni interface to account for a phonon transport channel whose dominance in ultrathin films was strongly suggested by our above derivations.

In analogy to recent investigations\cite{pude2020b,matt2022a}, we numerically solve the system of coupled differential eqs.~(\ref{eq1:diff2TM}) and (\ref{eq2:diff2TM}) using bulk parameters for the involved materials. In particular, we fix the EPC constant in Au to $g_\text{Au}=2\cdot10^{16}\,$W/(m$^3$\,K)\cite{hohl2000}. The thermal stress related to the heated subsystems drives coherent and incoherent spatiotemporal lattice strain which is calculated using a linear-chain model\cite{herz2012,schi2014,schi2021}. The average strain in the Au and Ni layers obtained by these calculations is compared to the UXRD data in Fig.~\ref{fig:UXRDdata}. We find a very satisfactory agreement of the numerical results with the experimental data for the various samples with different Ni and Au layer combinations. In particular, the initial compression of the cold Au lattice by the adjacent Ni layer due to its fast heating is well reproduced, evidencing the fast and efficient funneling of the excitation energy into the Ni phonons\cite{pude2020b,matt2022a}. The slow rise of the incoherent Au strain is matched in all samples simultaneously assuming a finite phonon interface conductance of $G_\text{ph} = 270$\,MW/(m$^2$\,K). In accord with the predictions by our m2TM summarized in Fig.~\ref{fig:results}(b), the build-up of the incoherent stress/strain in Au would be much slower when omitting the direct energy exchange between Ni and Au via phonon diffusion (not shown), in particular, for the samples with single-digit nanometer Au thickness.

We emphasize that the local internal mechanical stress inside a material is proportional to the local energy density of its subsystems\cite{Lindenberg2000,Nicoul2011,koc2017a,matt2021a}. That is, the incoherent lattice strain measured in our UXRD experiments directly quantifies the transient layer-averaged energy density. We thus discuss the energy transport processes from the perspective of our direct observable, i.e., the transient Au strain shown in Fig.~\ref{fig:UXRDdata}(c).

\begin{figure}[t]
  \centering
  %\begin{minipage}{\columnwidth}
    \includegraphics[width = \columnwidth]{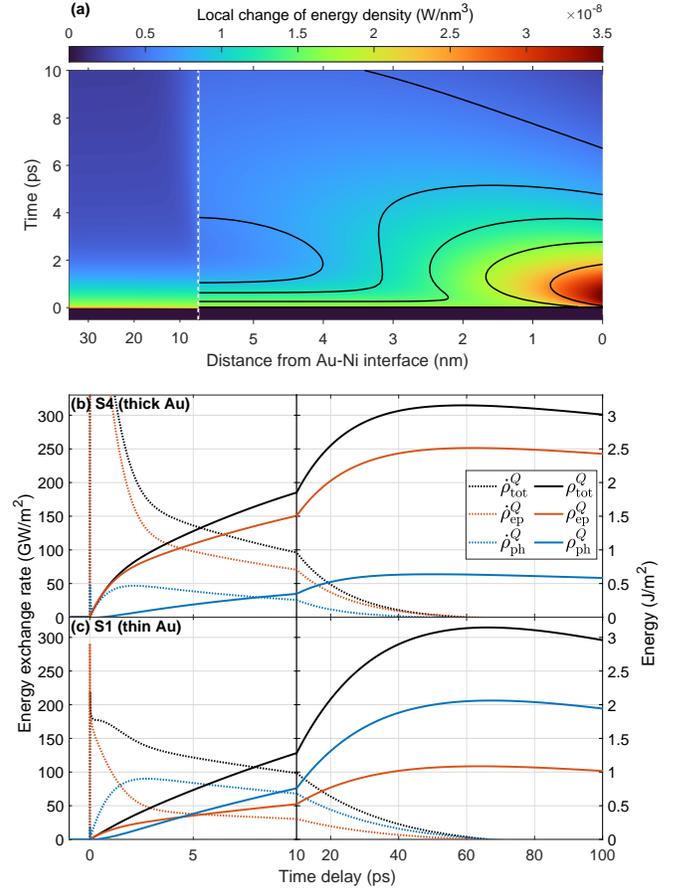}
    %\includegraphics[width = \columnwidth]{figures/fig4-1_axisBreak_reversed.eps}
  %\end{minipage}
  %\begin{minipage}{\columnwidth}
    %\includegraphics[width = 1\columnwidth]{figures/fig4bc.eps}
  %\end{minipage}%
  \caption{(a) Spatiotemporal energy density rates inside the Au phonon subsystem corresponding to the strain simulations shown in Fig.~\ref{fig:UXRDdata}(c) for the sample S1 with thinnest Au layer (contour lines) and S4 with the thickest Au layer (false-color plot). The interface to the Ni layer is located at the right-hand side of the plot.
  (b)-(c) Time evolution of the layer-integrated energy density rates $\dot\rho^Q$ (dotted lines) and accumulated energy $\rho^Q$ per unit area (solid lines) for sample S4 and S1, respectively. The line colors encode the total energy (black) and its separation into EPC (red) and phonon transport channels (blue). While EPC dictates the energy flow into thick Au, phonon transport dominates after a few picoseconds in thin Au.}
  \label{fig:energyFlux}
\end{figure}

In the following we dissect the d2TM given by eqs.~(\ref{eq1:diff2TM}) and (\ref{eq2:diff2TM}) thereby showing how to individually quantify the energy transport within the heterostructure by electrons and phonons. In essence, the left side of eq.~(\ref{eq1:diff2TM}) and (\ref{eq2:diff2TM}) represents the total time-dependent rate of local energy density $\dot\rho^Q_\text{tot}$ in the electron and phonon subsystems, respectively. 
To illustrate the two channels that funnel energy into the Au phonon subsystem we depict the total energy density rate $\dot\rho^Q_\text{tot,ph}=\dot\rho^Q_\text{ph} +\dot\rho^Q_\text{ep}$ corresponding to the calculation for sample S4 (34.4\,nm Au layer) in the false-color plot in Fig.~\ref{fig:energyFlux}(a). 
Note the broken abscissa scaling in order to highlight the region near the Au-Ni interface. Two main sources of Au phonon energy increase can be identified: (i) a spatially homogeneous energy input within the first picosecond due to EPC via the homogeneously distributed hot electrons in Au and (ii) energy entering inhomogeneously from the Ni layer (right-hand border of plot) by phonon diffusion. The contour lines superimposed on the false-color plot correspond to the rate of energy density transported into the 5.6\,nm thick Au layer of S1. Clearly, the local energy input into the Au phonons is nearly independent of the Au thickness. As already mentioned above, the main difference is that the thick film offers much more volume for energy transfer via EPC, which is proportional $d_\text{Au}$ (cf. eq.~(\ref{eq:cond})). Of course, this only holds up to roughly the electron-phonon diffusion length $L^\text{e}_\text{diff}$ mentioned above.
%The ultrashort laser excitation exclusively deposits energy density inside the system of Au and Ni conduction electrons which very quickly thermalize with the Ni lattice\cite{pude2020b,matt2022a}. Subsequently, t
%The local energy density in the Au lattice is changed due to either phonon diffusion from the hot Ni lattice ($\dot\rho^Q_\text{ph}$) or due to EPC ($\dot\rho^Q_\text{ep}$).

While Fig.~\ref{fig:energyFlux}(a) illustrates the energy flowing into the Au phonon system with full spatial resolution, we plot the spatially integrated quantities as dashed lines in Fig.~\ref{fig:energyFlux}b) for the thickest and in Fig.~\ref{fig:energyFlux}(c) for the thinnest Au film. Here we differentiate the two channels from eq.~\eqref{eq2:diff2TM} by color: Initially the EPC channel ($\dot\rho^Q_\text{ep}$, dashed red) dominates in both samples. In case of the thinner Au film (panel (c)), the phonon transport rate ($\dot\rho^Q_\text{ph}$, dashed blue) across the Au/Ni interface becomes dominant after about 2\,ps. Beyond 5\,ps more than half of the energy accumulated in the Au phonons was received via this channel (solid blue line). For the thicker Au film, however, the EPC is always the dominant channel (solid red line).
For the thinnest film S1 (Fig.~\ref{fig:energyFlux}(c)) we find that at the moment of maximum total energy arrived in the Au phonons, 2/3 of the energy density via phonon diffusion and only 1/3 via EPC. This phonon-dominated thermal transport is a remarkable observation given that two elemental metals are considered where thermal transport via phonons is usually at least an order of magnitude weaker than via conduction electrons\cite{cahi2014,wang2016}.

The results of the above analysis on all investigated samples is summarized in Fig.~\ref{fig:results}(d) as green triangles. Clearly, the quantitatively much more reliable d2TM verifies the conclusions drawn from the conceptually simpler m2TM, namely, that the energy exchange in our Au-Ni bilayers is dominated by phonon diffusion if the Au thickness falls below approx.~10\,nm.

\section{Conclusion}

We have investigated Au-Ni heterostructures with Au layer thicknesses ranging from 5.6 to 34\,nm excited by ultrashort laser puses. The modeling of UXRD experiments via a simplified (m2TM) and a more complex and quantitative two-temperature model (d2TM) (Fig.~\ref{fig:results}(d)) consistently reveals that energy transport from the hot Ni into the cold Au phonon system is dominated by phonon diffusion across the Au/Ni interface, provided the Au layer's thickness is on the order of the phonon diffusion length of approx.~10\,nm.
The main ingredient to realize dominant phonon heat transport in the discussed nanoscale heterostructures is the combination of the weak EPC in Au and the strong EPC in Ni. As a consequence, the energetic conduction electrons rapidly transfer the excess energy to the Ni phonons. Shortly after laser excitation, this gives rise to a large phonon temperature gradient causing a large energy flow across the Au-Ni interface via phonons. These processes at the interface occur independent of the Au layer thickness (see comparison in Fig.~\ref{fig:energyFlux}a)). When the Au thickness drops below the electron diffusion length of approx.~125\,nm, the energy transfer to Au phonons via EPC decreases linearly. For thicknesses on the order of the phonon diffusion length, the direct phonon transport prevails over the electronic channel. 
The strong EPC in Ni remains crucial during the energy transfer to the Au lattice, because any excess energy in the electron system migrates preferentially to Ni phonons, thus stabilizing the phonon temperature gradient across the Au/Ni interface.

We conclude that at the nanoscale, phonon thermal transport cannot be neglected even for noble metals if they are in contact with another metal with large EPC. We believe that this surprising fact will be useful in understanding and tailoring energy transport in various systems ranging from plasmonic heterostructures with applications in chemistry to (magnetic) data-processing and storage in nanoscale device structures.

%. We note the strong EPC in Ni was only used to efficiently  create a very large spatial gradient in the phonon temperature. The dominant phonon n efficital heterostructures with very different EPC strengths must not be neglected since the ????????? phonon diffusion length may be greatly enhanced due to the potentially long e-ph equilibration time scales in such composite materials.
\begin{acknowledgments}
This work was funded by the Deutsche Forschungsgemeinschaft (DFG, German Research Foundation) via BA
2281/11-1. We acknowledge the characterization
of the heterostructures via XRR at the XPP-KMC3 synchrotron radiation
beamline D13.2 at the BESSY II electron storage
ring operated by the Helmholtz-Zentrum Berlin.
\end{acknowledgments}

\bibliography{references.bib}
\end{document}